\documentclass[prd,aps,english]{revtex4}
\usepackage{epsfig}
\usepackage{graphicx,amsmath}

\usepackage{slashed}

\newcommand{\Li}{\mbox{Li}_2}

\newcommand\be{\begin{equation}}
\newcommand\ee{\end{equation}}
\newcommand\ba{\begin{eqnarray}}
\newcommand\ea{\end{eqnarray}}
\newcommand{\br}[1]{\left( #1 \right)}

\newcommand{\brs}[1]{\left[ #1 \right]}

\newcommand{\brm}[1]{\left| #1 \right|}
\newcommand\nn{\nonumber}

\def\Li#1#2{{\mathrm{Li}}_{#1}\left(#2\right)}

\newcommand{\GeV}{\mbox{GeV}}

\newcommand{\M}{{\cal M}}	

\renewcommand{\Re}{\mbox{Re}}

\newcommand{\EoM}{r}

\begin{document}
\title{Radiative corrections for electron proton elastic scattering taking into account high orders and hard photon emission}
\author{E.~A.~Kuraev}
\email{kuraev@theor.jinr.ru}
\affiliation{JINR-BLTP, 141980 Dubna, Moscow region, Russian Federation}

\author{A.~I.~Ahmadov}
\email{ahmadov@theor.jinr.ru}
\affiliation{JINR-BLTP, 141980 Dubna, Moscow region, Russian Federation}
\affiliation{Institute of Physics, Azerbaijan National Academy of Sciences, Baku, Azerbaijan}

\author{Yu.~M.~Bystritskiy}
\email{bystr@theor.jinr.ru}
\affiliation{JINR-BLTP, 141980 Dubna, Moscow region, Russian Federation}

\author{E.~Tomasi-Gustafsson}
\email{etomasi@cea.fr}
\affiliation{\it CEA,IRFU,SPhN, Saclay, 91191 Gif-sur-Yvette Cedex, France, and \\
CNRS/IN2P3, Institut de Physique Nucl\'eaire, UMR 8608, 91405 Orsay, France}

\date{\today}

\begin{abstract}
We investigate the effect of high order radiative corrections in unpolarized electron proton elastic scattering and compare with the calculations at lowest order, which are usually applied to experimental data.  Particular attention is devoted to the $\epsilon$ dependence of radiative corrections, which is directly related to the electric proton form factor. We consider in particular the effects of the interference terms for soft and hard photon emission. Both quadratic amplitude describing the collinear emission along the
scattered electron as well as the interference with the amplitudes of emission from the initial
electron and the emission from protons are important in leading and next to leading approximation and they may compensate in particular kinematical conditions.

\end{abstract}

\maketitle
\section{Introduction}
The field of electromagnetic nucleon form factors (FFs) gathers intense activity, due in particular, to new experimental opportunities which allow to extend measurements at large momentum transfer and/or to achieve larger precision. In particular the possibility to apply the recoil polarization method suggested by  \cite{Akhiezer:1968ek,Akhiezer:1974em} allowed a measurement of the electric to magnetic  FF ratio up to a value of the momentum transfer squared of $Q^2=8.5~\GeV^2$ by the GEp collaboration at the Jefferson Laboratory \cite{Puckett:2011xg} and Refs. therein. These results substantially differ from the traditional measurements based on the Rosenbluth method \cite{Andivahis:1994rq,Christy:2004rc,Qattan:2004ht,Perdrisat:2006hj}.

In Ref. \cite{Bystritskiy:2007hw} it was shown that the
radiative corrections (RC) which correspond to the emission of photons from
the initial electron can lead these two sets of experimental data into agreement
for the transfer momentum range up to $Q^2 \leq 4~\GeV^2/c^2$.
RC change essentially the unpolarized cross section and are usually applied at first order to the experimental data, according to the classical paper of Mo and Tsai \cite{Mo:1968cg}, more recently revised by Maximon and Tjon \cite{Maximon:2000hm}: the applied RC are sizable (up to 40\%) and strongly depend on the relevant variables, the momentum transfer squared $Q^2$ and the linear polarization of the virtual photon, $\epsilon$. It was shown in the leading-logarithmical
approximation (LLA) \cite{Baier:1980kx}  that RC decrease the slope of $\epsilon$-dependence of the
reduced cross section $\sigma_{red}$ in the Rosenbluth method), which is directly related to the electric form factor, $G_E$ (see Fig.~4 in \cite{Bystritskiy:2007hw}).
Non leading terms are taken into account as a $K$ factor which is of the order of unity. Among these contributions, in Ref. \cite{Bystritskiy:2007hw} it was shown that the hard two-photon exchange amplitude is small (does not exceed $1-2\%$). The charge-odd interference terms
which are related to virtual and soft real proton emission from
electron and proton lines were not taken into account.
These last terms were calculated within the first order of perturbation theory in Ref. \cite{Maximon:2000hm}.

In Ref. \cite{Bystritskiy:2007hw}, as well as in Ref. \cite{Maximon:2000hm} hard photon emission was not considered. In particular the charge-odd interference of the
hard photon emitted from the final electron and from proton line was left out
of consideration. The aim of this work is to calculate this contribution and to study not only its size but also its dependence on the relevant kinematical variables.

Using the same arguments as in \cite{Maximon:2000hm}, the dependence of the matrix element on the proton form factor is assumed to be smooth. We consider the  proton as a point-like particle. This approximation turns out to be quite good when calculating the relevant corrections which are included in the $K$-factor and which are expressed as the ratio of cross sections.  It is justified by the fact that in this ratio the dependence over the form factors mostly cancels.

In Refs. \cite{TomasiGustafsson:2006pa,Bystritskiy:2007hw}, it was shown that Lepton Structure Function (LSF), which takes into account high order radiative corrections (RC) in leading logarithm approximation, gives different RC, changing the size of the observables as well as their dependence on the relevant kinematical variables.

The main contribution to LSF, in $ep$ elastic scattering, contains the terms in $L=\log ( Q^2/m^2_e)$ ($m_e$ is the electron mass).  One can see that, already at $Q^2\sim 1~\GeV^2$, such terms become large ($L\sim 15$) partially compensating the factor $\alpha/\pi$ which accompanies the emission of an additional photon.

In Ref. \cite{Bystritskiy:2007hw} it was shown that LSF gives the largest contribution to the slope of the reduced cross section as a function of $\epsilon$, that the slope depends on the inelasticity cut and that different calculations give different slopes.
In Ref. \cite{Bystritskiy:2007hw} charge-odd contributions were neglected as it is known that a (partial) compensation exist, when taking into account hard photon contribution \cite{Kuraev:1977rp}.

In order to extract precise information on the hadron structure, it is necessary to carefully correct the electron block for the emitted photons. This is especially true when the experiment is not fully exclusive, but also in this case, radiative corrections have to be applied within the acceptance and the resolution of the detection.

The purpose of this paper is to complete the calculation of Ref. \cite{Bystritskiy:2007hw} adding the hard photon contribution. Moreover, the effects on polarization observables, which can be calculated in a straightforward way in the frame of SF method, are studied.

In Section~\ref{sec.Formalism}, the LSF formalism is briefly recalled, and the relevant expressions are given and discussed. In Section~\ref{sec.HardPhotonEmission} the main results will be presented. The different terms will be compared in first and higher order calculations. In Conclusions we do a brief summary, comment the contribution of inelastic channels, and stress the importance of including high order RC in the codes for the experimental analysis.

\section{Formalism}
\label{sec.Formalism}

Let us consider the process of elastic electron-proton scattering
(see Fig.~\ref{FigBorn}):
\begin{figure}
    \includegraphics[width=0.35\textwidth]{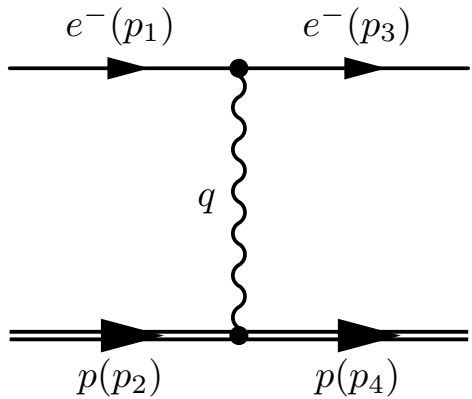}
    \caption{\label{FigBorn}
        Amplitude of electron-proton scattering in Born approximation.
    }
\end{figure}
\ba
    e^-\br{p_1} + p\br{p_2} \to e^-\br{p_3} + p\br{p_4},
\ea
where $p_1^2 = p_3^2 = m_e^2$ and $p_2^2=p_4^2=M^2$.
The matrix element for this process in Born approximation
and for point-like proton formfactor
(i.e. $F_1=1$ and $F_2=0$) is:
\ba
    \M_B = \frac{e^2}{Q^2}
    \brs{\bar u\br{p_3} \gamma^\mu u\br{p_1}}
    \brs{\bar u\br{p_2} \gamma_\mu u\br{p_4}},
\ea
with $e^2 = 4\pi\alpha$, ($\alpha\approx 1/137$ is the fine structure constant)
and $Q^2 = -q^2$, where $q=p_1-p_3=p_4-p_2$ is the transferred four-momentum.

The differential cross section in one photon approximation can be expressed as a function of two kinematical variables,
$Q^2$ and $\epsilon$, the four momentum $Q^2$ and the polarization $\epsilon$ of the exchanged virtual photon, in the form:
\be
\frac{d\sigma}{d\Omega}
(Q^2,\epsilon)=\frac{\sigma_M}
{\epsilon
(1+\tau)}\sigma_{red}(Q^2,\epsilon),
\qquad
\sigma_{red}(Q^2,\epsilon)={\tau} \,G_M^2(Q^2)+\epsilon \, G_E^2(Q^2),~
\label{eq:eqphi}
\ee
with
\be
\sigma_M=\frac{Z^2\alpha^2\cos^2(\theta_e/2)}{4E^2\sin^4(\theta_e/2)},
\qquad
Q^2=-4 E E'\sin^2\displaystyle\frac{\theta_e}{2},
\qquad
\tau=\frac{Q^2}{4M^2},
\qquad
\frac{1}{\epsilon }=1+2(1+\tau)\tan^2\displaystyle\frac{\theta_e}{2}.
\label{eq:eqtau}
\ee
where $\sigma_M$ is the Mott's cross section for electron scattering on point-like particles, and the nucleon structure is described by the form factors, $G_E$ and $G_M$. The kinematical variables are expressed as a function of the incident(final) electron energy $E$ ($E'$), the electron scattering angle $\theta_e$. Eqs. (\ref{eq:eqphi},\ref{eq:eqtau}) hold for elastic electron scattering on spin 1/2 hadron, with appropriate values of the mass and the charge of the hadron $M$, $Z$.

It is known \cite{Kuraev:1988xn} that the process of emission of hard photons by initial and scattered electrons plays a crucial role, which results in the presence of the radiative tail in the distribution on the scattered electron energy. The LSF approach extends the traditional calculation of radiative corrections \cite{Mo:1968cg}, taking into account the contributions of higher orders of perturbation theory and the role of initial state photon emission. The cross section can be  expressed in terms of LSF of the initial electron and of the fragmentation function of the scattered electron energy fraction:
\be
d\sigma^{LSF}(Q^2,\epsilon)=
\int_{z_0}^1 dz {\cal D}(z,\beta)
d\tilde\sigma(Q^2_z,\epsilon_z)\left (1+\frac{\alpha}{\pi}K \right ),
\qquad
\mbox{with}
\qquad
d\tilde\sigma(Q^2_z,\epsilon_z)= \frac{d\sigma^B(Q^2_z,\epsilon_z)}{|1-\Pi(Q^2_z)|^2},
\label{eq:eqy}
\ee
where $d\tilde\sigma(Q^2_z,\epsilon_z)$,  is the Born cross section corrected by the vacuum polarization, calculated for a kinematics shifted by $z$. The $z$-dependent kinematical variables, taking into account the change of the electron four momentum, due to photon emission,
$Q^2_z$, $\epsilon_z$ are calculated from the corresponding ones (Eq. (
\ref{eq:eqtau})), replacing the initial electron energy $E$ by $zE$, which is the energy fraction carried by the electron after emission of one or more collinear photons.

We used for simplicity the notation $d\sigma$ for the double differential cross section:
$d\sigma^{LSF,B}= (d\sigma^{LSF,B}/{d\Omega})$, for Born approximation $(B)$ and radiatively corrected $(LSF)$.
The lower limit of integration, $z_0$, is related to the 'inelasticity' cut, $c$, used to select the elastic data, and corresponds to the maximum energy of the soft photon, which escapes the detection:
\be
z_0=\frac{c}{\rho -c(\rho-1)},
\label{eq:eqz}
\ee
where  $\rho$ is the recoil factor $\rho=1+(E/M)(1-c_e)$ and $y=1/\rho $ is the fraction of incident energy carried by the scattered electron. In terms of $\rho$, one can write $Q^2={2E^2(1-c_e)}/{\rho}$,
with the notation $c_e=\cos\theta_e$.

In Eq. (\ref{eq:eqy}) the main role is played by the non singlet LSF:
\be
{\cal D}(z,\beta)=\frac{\beta}{2}\brs{\br{1+\frac{3}{8}\beta}(1-z)^{\frac{\beta}{2}-1}-
\frac{1}{2}(1+z)} \br{1+O(\beta)},
\label{eq:eq6}
\ee
\be
    \beta=\frac{2\alpha}{\pi}(L-1), \qquad L=\ln\frac{Q^2}{m_e^2},
\label{eq:eq6a}
\ee
$m_e$ is the electron mass. Particularly important is the quantity $L$, called, large logarithm, which is responsible for the large size of the term related to the LSF correction.

The integration in Eq. (\ref{eq:eqy}) requires a careful treatment, as ${\cal D}(z)$ has a singularity for $z=1$. So the integration of any function $\Phi$ gives (see Appendix A in \cite{Bystritskiy:2007hw}):
\ba
{\cal I}&=&\int_{z_0}^1D(x)\Phi(x) dx= \nn\\
&=&\Phi(1)\left [ 1-\displaystyle\frac{\beta}{4}\left (2\ln \displaystyle\frac{1}{1-z_0}-z_0-\displaystyle\frac{z_0^2}{2}\right )\right ]
+\displaystyle\frac{\beta}{4}\int_{z_0}^1 dx\displaystyle\frac{1+x^2}{1-x}\left [\Phi(x)-\Phi(1)\right ]+{\cal O}(\beta^2).
\label{eq:eqa1}
\ea

The factor $1+({\alpha}/{\pi })K$ in Eq. (\ref{eq:eqy}) has been calculated in detail for $ep$ elastic scattering in Ref. \cite{Bystritskiy:2007hw,Maximon:2000hm} where the term $K$ is the sum of three contributions:
\be
K=K_e+K_p+K_{b}.
\label{eq:eqkfac}
\ee
$K_e$ is related to non leading contributions arising from the pure electron block  and can be written as \cite{Kuraev:1985hb,Kuraev:1988xn}:
\be K_e= -\displaystyle\frac{\pi^2}{6} -\displaystyle\frac{1}{2} -\displaystyle\frac{1}{2}\ln^2\rho+\Li{2}{\cos^2\theta_e/2},
\qquad
\Li{2}{z}=-\int_0^z \displaystyle\frac{dx}{x} \ln(1-x).
\label{eq:eqll}
\ee
The second term, $K_p$,  concerns the emission from the proton block. The emission of virtual and soft photons by the proton is not associated with large logarithm, $L$, therefore the whole proton contribution can be included as a $K_p$ factor:
\begin{eqnarray}
K_p&=&\displaystyle\frac{Z^2}{\beta}\left \{ -\displaystyle\frac{1}{2}\ln^2x-\ln x\ln [4(1+\tau)]
+\ln x - \right .
\nonumber \\
&&\left . -(\ln x-\beta)\ln\left [ \displaystyle\frac{M^2}{4E^2(1-c)^2}\right ]+\beta -\Li{2}{1-\displaystyle\frac{1}{x^2}}+2\,\Li{2}{-\displaystyle\frac{1}{x}}+
\displaystyle\frac{\pi^2}{6} \right\},
\label{eq:eqkp}
\end{eqnarray}
with $x=(\sqrt{1+\tau}+\sqrt{\tau})^2$, $\beta=\sqrt{1-M^2/E'^2}$ and $E'=E(1-1/\rho)+M$ are the scattered proton velocity and energy.
The contribution of $K_p$ to the $K$ factor is of the order of -0.2\% for $c=0.99$, $E=21.5~\GeV$, $Q^2=31.3~\GeV^2$ \cite{Maximon:2000hm}, and it is almost constant in $\epsilon$.

Lastly, $K_{b}$ represents the interference of electron and proton emission. More precisely the interference between the two virtual photon exchange amplitude and the Born amplitude as well as the relevant part of the soft photon emission i.e., the interference between the electron and proton soft photon emission, may be both included in the term $K_{b}$. These effects are not enhanced by large logarithm (characteristic of LSF) and can be considered among the non-leading contribution, which represents an $\epsilon$-independent quantity of the order of unity, including all the non-leading terms, as two photon exchange and soft photon emission.

Here we will consider two additional contributions to the $K$-factor, $K_{h}$ from hard photon emission and the charge odd contribution from the interference between  electron and proton emission, $K_{o}$
\ba
K=K_e+K_p+K_{b}+K_{h}+K_{o}.
\label{eq:eqktotal}
\ea
In order to make comparison with existing calculations of RC, it is convenient to express the corrections calculated with the LSF method, $\delta $ in the form:
\be
d\sigma^{LSF}(Q^2,\epsilon)=d\sigma^B(Q^2,\epsilon)(1+\delta),
\label{eq:csa}
\ee
where
\ba
1+\delta&=&
\frac{1}{|1-\Pi(Q^2)|^2}
\left \{
1+\displaystyle\frac{\alpha}{2\pi}(L-1)
\left [
-\left (2\ln\left (\displaystyle\frac{1}{1-z_0}\right )-z_0-\displaystyle\frac{z_0^2}{2}
\right )+
\right .\right .
\nn\\
&+&
\left .
\int_{z_0}^1 dz\displaystyle\frac{1+z^2}{1-z}
\left .
\left(
\displaystyle\frac
{d\tilde\sigma^B(Q^2_z,\epsilon_z)}
{ d\tilde\sigma^B(Q^2,\epsilon)}-1\right ) \right ]
+\displaystyle\frac{\alpha}{\pi} K.\right \}
\label{eq:cs}
\ea
Let us compare these different terms with the corresponding calculation from Ref. \cite{Maximon:2000hm}, which has removed or softened some drastic approximations previously used in \cite{Mo:1968cg}. The interference between the box and the Born diagram was included (partially within the soft photon approximation) as:
\be
\delta^{box}=\frac{2\alpha }{\pi}Z\left \{ -\ln\rho\ln
\left [\frac{-q^2x}{(2\rho\Delta E)^2} \right ]
+ \Li{2}{1-\frac{\rho }{x}}
- \Li{2}{1-\frac{1}{\rho x}}\right \},
\label{eq:MTbox}
\ee
where $\Delta E= E'(1-c)$ is the maximum energy of the soft photon, allowed by the experimental set-up.
The radiation from the electron, in the leading order approximation, including vacuum polarization, was expressed as
\be
\delta^{el}=\frac{\alpha }{\pi}\left \{ \frac{13}{6}L
 - \frac{28}{9} -( L -1 ) \ln\left [\frac {4 EE'}{(2\rho \Delta E)^2}\right ]
-\frac{1}{2}\ln^2\rho + \Li{2}{\cos^2\frac{\theta_e}{2}}
-\frac{\pi^2}{6}\right \}.
\label{eq:MTel}
\ee


In the LSF calculation, the main contribution comes from those terms, which include higher order corrections, whereas all the terms which do not contain large logarithm are expected to be suppressed and included in the $K$-factor.

\begin{figure}
\begin{center}
\includegraphics[width=12cm]{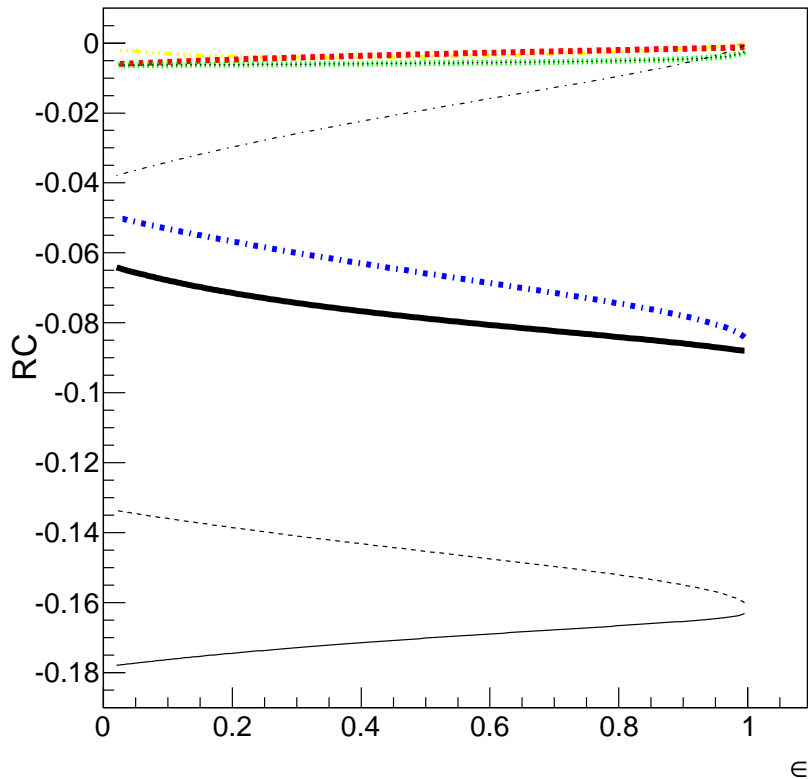}
\caption{\label{Fig:rc1GeV}(Color online)Calculation of different radiative corrections for $Q^2=1~\GeV^2$.
Thick lines correspond to LSF calculation, thin black lines correspond to MT \protect\cite{Maximon:2000hm} calculation. Total correction (black, solid lines), electron emission (red ($K_e$) and black dashed lines), proton emission (green and black dotted lines), Structure function (LL) (blue dash-dotted line), box MT (black, thin, dash-dotted line), two-photon contribution from \protect\cite{Bystritskiy:2007hw}(yellow triple dotted-dashed line)  }
\end{center}
\end{figure}

In Fig. \ref{Fig:rc1GeV}  the results for the calculation of different radiative corrections for $Q^2=1~\GeV^2$ and $\Delta E=0.03 E'$ are shown. Thick lines correspond to the LSF method and thin lines correspond to MT \cite{Maximon:2000hm}. The solid line is the sum of the different terms. A large difference can be already seen at such small value of $Q^2$, both in the values and in the slope of the cross section.  The numerical values of the radiative corrections are larger at larger $Q^2$ and the effect of higher orders becomes more sizable. The final RC factor depends on both $\epsilon$ and $Q^2$.

Let us compare the different terms. The correction from the proton (dash-dotted line) is basically the same, in both calculations (Eq. \ref{eq:eqkp}). It is small and $\epsilon$ independent. For both methods, the largest contribution is due to the radiation from the electron. In the LSF method, the main correction is due to the electron radiation (dash-dotted line), whereas the corrections from the electron which do not contain large logarithm and are  calculated in the $K_e$ factor, which is small (dashed line), with a small $\epsilon$ dependence. The electron emission from the MT calculation, Eq. (\ref{eq:MTel}) is shown as a thin, dashed line, and corresponds to the largest contribution to RC. In the LSF method, only the emission from the initial electron is taken into account. For the  final electron emission, it has been assumed that the full energy is detected (for example, if the electron is detected in a calorimeter) or that the electron energy is not measured at all: in these cases, due to the properties of LSF, the contribution for final emission is unity. This explain the difference of about a factor of two in the total contribution.

For the LSF calculation, $K_{b}$ has been calculated as in \cite{Bystritskiy:2007hw}, with integration over loop momenta in the region when two photons are hard ($\brm{q_{1,2}^2} \approx Q^2/2$). It was shown that this term is small, and with small $\epsilon$ dependence.

In the MT calculation, the interference between the box and the Born diagram (thin, black dash-dotted line) has a positive slope, and a large  $\epsilon$ dependence (Eq. (\ref{eq:MTbox})). Therefore it is this term which is responsible for the fact that the slopes of the final corrections as a function of $\epsilon$ (thin and thick solid lines) have opposite signs in the LSF and MT calculations. When applied to the experimental cross sections, this will be reflected in a change of slope of the reduced cross section, as a function of $\epsilon$, and the electromagnetic FFs extracted from the Rosenbluth method will be different. In Ref. \cite{Bystritskiy:2007hw} it has been shown that the LSF corrections could bring into agreement the FFs extracted by the Rosenbluth and the polarization methods.

To summarize, the main difference between the size of RC from the two calculations should be attributed to the fact that in the present application of the LSF method, the partition function of the final particle is taken as unity, which is the case in an experiment where one can not separate events corresponding to an electron and to an electron and a photon with the same total energy. The difference between the slopes of the corrections to the cross section as a function of $\epsilon$  depends on the ansatz used to include the two photon exchange mechanism.

\section{Calculation of Hard photon emission}
\label{sec.HardPhotonEmission}

\begin{figure}
    \includegraphics[width=0.7\textwidth]{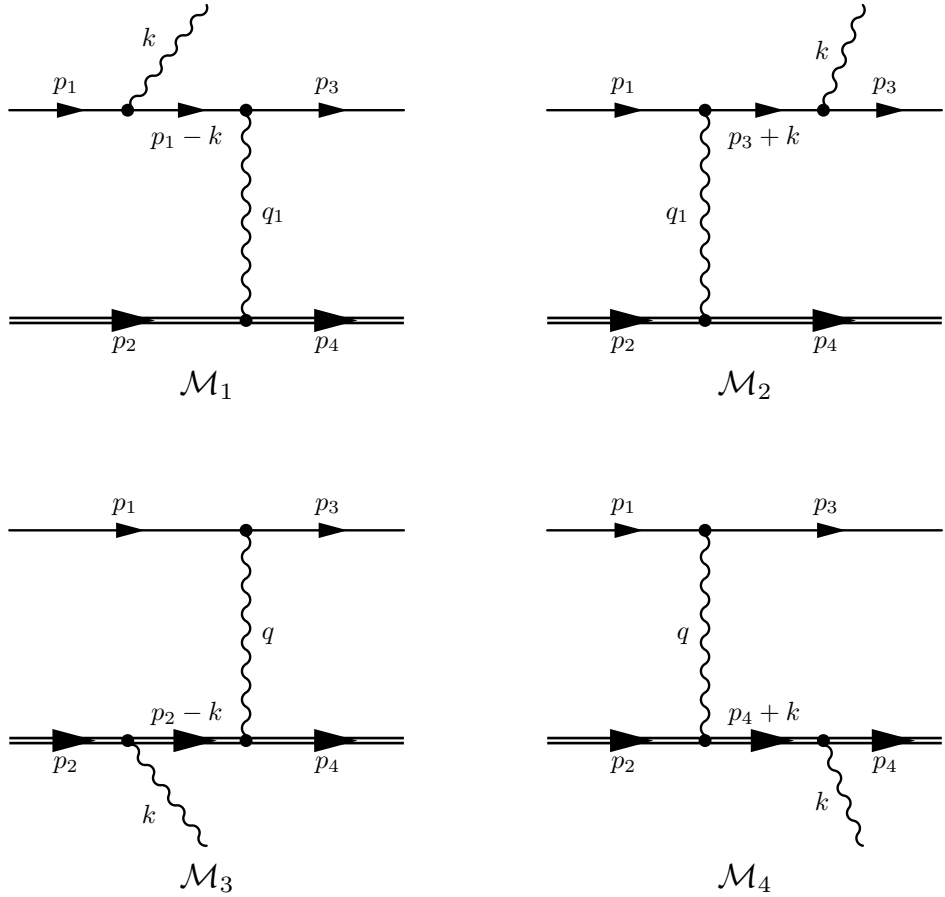}
    \caption{\label{FigHardPhoton}
        Diagrams contributing  to the amplitude of  hard photon emission in
        electron-proton scattering.
    }
\end{figure}

Let us consider the process of hard photon emission with momentum $k=\br{\omega, \bf k}$ in electron-proton scattering
(see Fig.~\ref{FigHardPhoton}):
\ba
e(p_1)+p(p_2) \to e(p_3)+p(p_4)+\gamma(k),
\label{eq:eqpgamma}
\ea
The cross section is given in Ref. \cite{Maximon:2000hm} in soft photon approximation:
\ba
d\sigma_B=-\frac{\alpha}{4\pi^2}d\sigma_0\int'\frac{d^3 k}{\omega}\br{\frac{p_3}{p_3k}-\frac{p_1}{p_1k}-Z\frac{p_4}{p_4k}+Z\frac{p_2}{p_2k}}^2.
\ea
Here the sign prime meant that the energy of the emitted soft photon do not exceed some small
quantity $\omega=\sqrt{k^2+\lambda^2}<\Delta E \ll E$ with $E$ is the energy in laboratory frame
of the initial electron. $Z$ is the charge of the fermion target in units of the electron charge $e$.
The value $Z=+1$ corresponds to the positively charged proton.
For this case the energy of the scattered electron is
\ba
E_{max}=\frac{ME}{M+E(1-c_e)}=E y_m, \qquad y_m=\frac{\EoM}{\EoM +1-c_e}, \qquad \EoM=\frac{M}{E},
\qquad c_e = \cos\theta_e,
\ea
where $\theta_e$ is the angle between the directions of initial and the scattered electrons. For the case of
emission of hard photon one have
\ba
E_3=\frac{ME-\omega(M+E(1-a_1))}{M+E(1-c_e)-\omega(1-c_3)}, \qquad c_3=\cos\theta_3, \qquad a_1=\cos\theta_1,
\ea
with $\theta_{1(3)}$-the angle between the directions of photon and initial (scattered) electrons.
In general case we have $E'+\omega<E_{max}$.
For collinear kinematics of photon emission $\theta_3=0$ we have
\ba
y_m=y+x, \qquad y=\frac{E'}{E}, \qquad x=\frac{\omega}{E}.
\ea
So for an experimental set-up where the scattered electron is detected in a calorimeter
the probability of hard photon emission by the scattered electron and by protons has to be taken into account,
as well as the interference of the corresponding amplitudes. We define the emission of a real photon with energy exceeding $\Delta E$ as the emission of a hard photon.

For the differential cross section we obtain
\ba
\frac{d\sigma}{d c_e}=\frac{d\sigma_B}{d c_e}(1+\delta)+\frac{d\sigma_h}{d c_e}.
\ea
A more elaborated result for $\delta$ was obtained in Ref. \cite{Maximon:2000hm}.

In the approximation of a point-like proton $F_1=1,F_2=0$, one finds:
\ba
\frac{d\sigma_B}{d c_e}=\frac{\pi\alpha^2 Z^2\cos^2(\theta_e/2)}{2E^2\sin^4(\theta_e/2)}y_m[1-2\tau\tan^2(\theta_e/2)], \qquad
q^2=-2E^2y_m z,
\qquad z = 1-c_e.
\ea
The contribution from the channel with hard photon emission
\ba
\frac{d\sigma_h}{dc_e}=\frac{(4\pi\alpha)^3Z^2}{16ME}\sum\brm{M_h}^2 \frac{d\Gamma_3}{dc_e},
\ea
will be considered below.
The matrix element of hard photon emission
has the form
$M_h=M_p+M_e$ with
\ba
M_p&=&-\frac{Z}{q^2}\bar{u}(p_4)O^N_\rho(e)u(p_2) \bar{u}(p_3)\gamma_\rho u(p_1); \nn \\
M_e&=&\frac{1}{q_1^2}\bar{u}(p_3)O^e_\sigma u(p_1) \bar{u}(p_4)\gamma_\sigma u(p_2); \nn \\
O^N_\rho(e)&=&O_{\rho\lambda} e_\lambda;\qquad
O^N_{\rho\lambda}=-\frac{1}{d_2}\gamma_\rho(\hat{p}_2-\hat{k}+M)\gamma_\lambda+\frac{1}{d_4}\gamma_\lambda(\hat{p}_4+\hat{k}+M)\gamma_\rho, \nn \\
O^e_\sigma(e)&=&O^e_{\sigma\lambda} e_\lambda;\qquad
O^e_{\sigma\lambda}=-\frac{1}{d_1}\gamma_\sigma(\hat{p}_1-\hat{k}+m)\gamma_\lambda+\frac{1}{d_3}\gamma_\lambda(\hat{p}_3+\hat{k}+m)\gamma_\sigma, \nn \\
&&d_1=2p_1k, \quad d_2=2p_2k; \quad d_3=2p_3k; \quad d_4=2p_4k; \quad q=p_1-p_3; \quad q_1=p_2-p_4.
\nn
\ea

In an experimental set-up with detection of the scattered electron, accompanied by a hard photon emitted within a narrow cone around
the direction of the scattered electron, only the terms with $1/d_3$, $m^2/d_3^2$ are relevant.
Using this prescription we obtain:
\ba
    R&=&E^2 \br{\sum \brm{M}^2}_{coll}
    =
    E^2\br{
        \brm{M_3}^2 + 2\Re\brs{M_1 M_3^* + M_2 M_3^* + M_3 M_4^*}
    }_{coll}
    =\nn\\
    &=&
    \frac{8 A \br{z_3 x^2 y - \frac{m^2}{E^2} y_m}}{\EoM\br{1-y_m}^2\br{x y z_3}^2}
    +
    \frac{4 y_m A}{\EoM \br{1-y_m}^2 x^2 z_3}
\nn\\
    &+&
    \frac{4 Z \EoM y_m A}{\br{1-y_m} x^2 y^2 z_3 (1-c_e)}
    -
    \frac{4 Z y_m^2 B}{\EoM \br{1-y_m} x^2 y z_3 (1-c_e)},
\ea
where we used the scalar products:
\ba
    &&
    \frac{q^2}{E^2} = -2y z,
    \qquad
    \frac{q_1^2}{E^2} = -2y \EoM(1-y_m),
    \qquad
    2\br{p_1 k}/E^2 = 2 x z,
    \nn\\
    &&
    2\br{p_1' k}/E^2 = 2 x y z_3,
    \qquad
    2\br{p_2' k}/E^2 = \frac{2 \EoM x}{y_m},
    \nn
\ea
and
\ba
    A = 2\EoM + z\br{1-\EoM-y_m},
    \qquad
    B = z^2 + \br{2-z}\EoM\br{\EoM + z}.
\ea
The phase volume of the final state can be written in the form
\ba
d\Gamma_3=\frac{1}{(2\pi)^5}\frac{d^3p_3}{2E_3}\frac{d^3p_4}{2E_4}\frac{d^3k}{2\omega}\delta^4(p_1+p_2-p_3-p_4-k)=
\frac{E^2 d c_e}{8(2\pi)^3}d\gamma,
\ea
with
\ba
d\gamma=y \, d y \,x \, d x\frac{d O_\gamma}{2\pi}\delta[\EoM(1-y-x)-y(1-c_e)-x(1-a_1)+y x(1-c_3)].
\ea
The angular phase volume of the photon can be written as:
\ba
\frac{dO_\gamma}{2\pi}=\frac{1}{\pi}\frac{d c_1 d c_3}{\sqrt{D(c_e,c_1,c_3)}}, \qquad D(c_e,c_1,c_3)=1-c_e^2-c_1^2-c_3^2+2c_e c_1 c_3,
\ea
where the condition $D(c_e,c_1,c_3)>0$ is implied. Imposing the experimental condition
\ba
1-\eta<c_3<1,
\ea
we write the phase volume as
\ba
\int\frac{d\Gamma_3}{E^2d c_e}=\frac{1}{64\pi^3}\int\limits_{1-\eta}^1d c_3\int\limits_{c_-(c_3)}^{c_+(c_3)}\frac{d c_1}{\pi\sqrt{D(c_e,c_1,c_3)}}\int\limits_\Delta^\Lambda
\frac{x[\EoM-x(\EoM+1-c_1)]}{[\EoM+1-c_e-x(1-c_3)]^2}d x, \nn \\
c_\pm(c_3)=c_e c_3\pm\sqrt{(1-c_e^2)(1-c_3^2)},
\ea
where
\ba
\Delta=\frac{\Delta E}{E}, \qquad \Lambda=\min\left [y_m=\frac{\EoM}{\EoM+1-c_e},\frac{\EoM}{\EoM +1-c_1}\right ].
\ea

In the collinear limit ($\eta \ll 1$) we can consider $c_1\approx c_e$ and the integral over $dc_1$ can be calculated explicitly using the formula:
\ba
    \int\limits_{c_-(c_3)}^{c_+(c_3)}\frac{d c_1}{\pi\sqrt{D(c_e,c_1,c_3)}} = 1,
\ea
and thus the phase volume takes the form:
\ba
\int\left.\frac{d\Gamma_3}{E^2d c_e}\right|_{coll}=\frac{1}{64\pi^3}\int\limits_{1-\eta}^1d c_3
\int\limits_\Delta^{y_m} dx
\frac{x y y_m}{\EoM}.\nn
\ea
And then using the integrals:
\ba
    \int\limits_{1-\eta}^1 \frac{dc_3}{1-\beta_3 c_3}
    &=&
    \ln\frac{2\eta E^2 y_m^2}{m^2} + 2\ln\br{1-\frac{x}{y_m}},
    \qquad
    \int\limits_{1-\eta}^1 \frac{dc_3 (m^2/E^2)}{\br{1-\beta_3 c_3}^2}
    =
    2 y^2,
    \nn\\
    \int\limits_{\Delta}^{y_m} \frac{dx}{x}
    &=&
    \ln\frac{y_m}{\Delta},
    \qquad
    \int\limits_{\Delta}^{y_m} \frac{dx}{x}
    \ln\br{1-\frac{x}{y_m}}
    =
    -\zeta_2,
    \nn
\ea
we get the following expression for the cross section of hard photon emission in the kinematics collinear to the final electron emission
\ba
\left.\frac{d\sigma}{d c_e}\right|_{coll}
&=&
\left.\frac{d\sigma}{d c_e}\right|_{33}+
\left.\frac{d\sigma}{d c_e}\right|_{13}+
\left.\frac{d\sigma}{d c_e}\right|_{23}+
\left.\frac{d\sigma}{d c_e}\right|_{34},
\\
\left.\frac{d\sigma}{d c_e}\right|_{33}&=&\frac{\alpha^3 Z^2 y_m A}{2 E^2 \EoM (1-c_e)^2}\br{\frac{1}{2}L-2\ln\frac{y_m}{\Delta}+\frac{1}{2}}, \nn \\
\left.\frac{d\sigma}{d c_e}\right|_{13}&=&\frac{\alpha^3Z^2 A}{4 E^2 \EoM (1-c_e)^2}\br{L(\ln\frac{y_m}{\Delta}-1)+2-\frac{1}{3}\pi^2}, \nn \\
\left.\frac{d\sigma}{d c_e}\right|_{23}&=&\frac{\alpha^3Z^3 y_m A}{4 E^2 (1-c_e)^2}\br{L\ln\frac{y_m}{\Delta}-\frac{1}{3}\pi^2}, \nn \\
\left.\frac{d\sigma}{d c_e}\right|_{34}&=&-\frac{\alpha^3Z^3 y_m^2 B}{4 E^2 \EoM(1-c_e)^2}\br{L \ln\frac{y_m}{\Delta}-\frac{1}{3}\pi^2},
\label{eq:hard}
\ea
where $L$ is the large logarithm
\ba
L=\ln\frac{2\eta y_m^2E^2}{m_e^2}.
\ea


\section{Cancellation between Hard photon and interference term}

\subsection{$K$-factors}
\label{SectionKFactors}

The expression for the cross section of hard photon emission
\ba
    d\sigma_h = \frac{1}{8s}
    2~\Re\brs{\br{\M_e}^+ \br{\M_p}}
    d\Gamma_3.
\ea
in a factorized form
\ba
    \frac{d\sigma}{dc_e}
    =
    \frac{d\sigma_B}{dc_e}
    +
    \frac{d\sigma_h}{dc_e}
    =
    \frac{d\sigma_B}{dc_e}
    \br{1+\delta_h},
\ea
where
\ba
    \delta_h = \br{\frac{d\sigma_B}{dc_e}}^{-1}\frac{d\sigma_h}{dc_e}
    = \frac{\alpha}{\pi} K_h.
    \label{KDef}
\ea

\begin{figure}
\includegraphics[width=12cm]{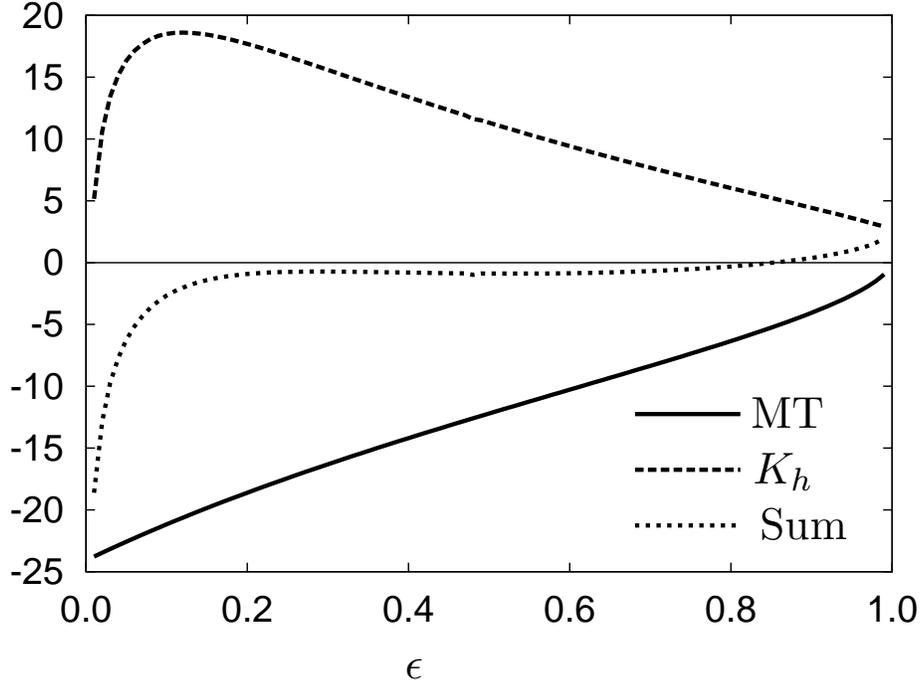}
\caption{\label{Fig1GeV2}
The $\epsilon$-dependence of $K$ factors : $K_h$ from  Eq.~(\ref{eq:hard}) (dashed line), $K_o$, from Ref. \protect\cite{Maximon:2000hm} (solid line), and their sum (dotted line) for $Q^2=1~\GeV^2$ and inelasticity cut $c=0.995$.
}
\end{figure}

\begin{figure}
\includegraphics[width=12cm]{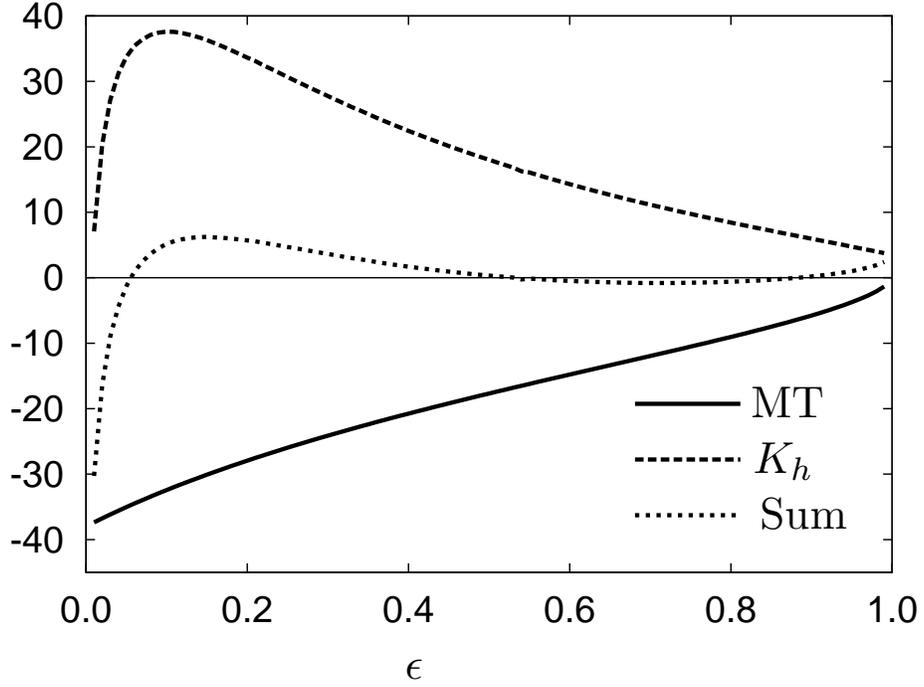}
\caption{\label{Fig3GeV2}
Same as Fig. \protect\ref{Fig1GeV2} but for $Q^2=3\GeV^2$.
}
\end{figure}

The results of the numerical evaluation are presented in
Figs.~\ref{Fig1GeV2}, \ref{Fig3GeV2}, for a value of inelasticity cut $c=0.995$
and $Q^2=1$, 3, and 5 GeV$^2$ respectively, as function of $\epsilon$.
The $K_h$-factor of hard photon emission
as calculated from Eqs.~(\ref{KDef}), (\ref{eq:hard}) (dashed line) is compared with the contribution for soft photon emission, from Ref. \cite{Maximon:2000hm}, Eq.~(5.2), 2nd line (solid line). The sum of these
two contributions (dotted line) shows an almost complete compensation of the
contributions of soft and hard photon emission
in the charge-odd interference, in all the $\epsilon$ region, starting from
$\epsilon\approx 0.2$.

\begin{figure}
\includegraphics[width=12cm]{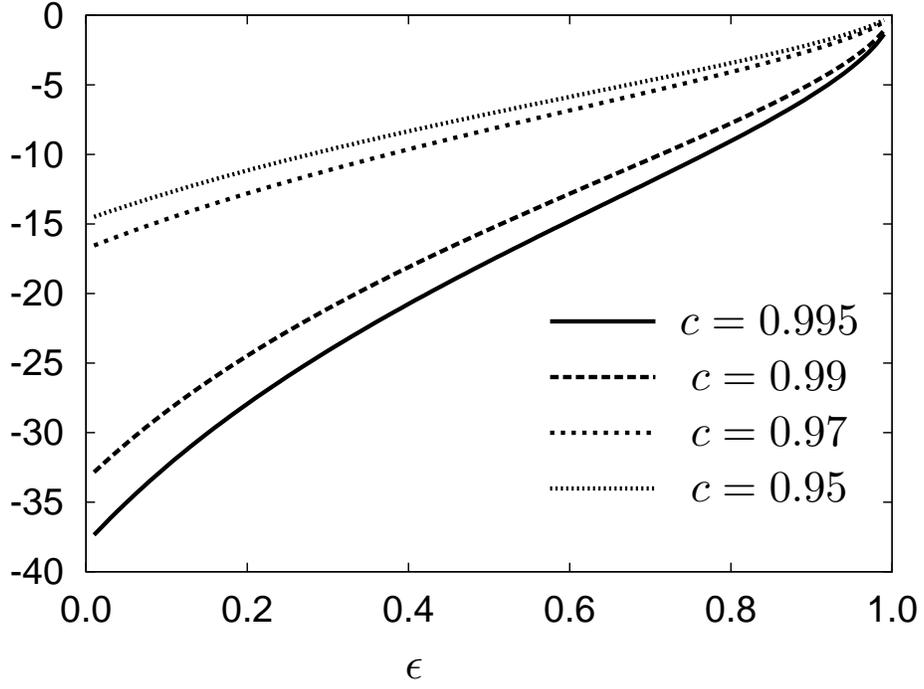}
\caption{\label{FigCutsMT}
$\epsilon$-dependence of virtual and soft real photon
charge-odd contributions $K_o$-factor (according to \cite{Maximon:2000hm})
for $Q^2=3~\GeV^2$ and for inelasticity cut $c$=0.95 (dotted line), $c$=0.97 (short dashed line), $c$=0.99 (dashed line) and $c$=0.995 (dashed line) .
}
\end{figure}

\begin{figure}
\includegraphics[width=12cm]{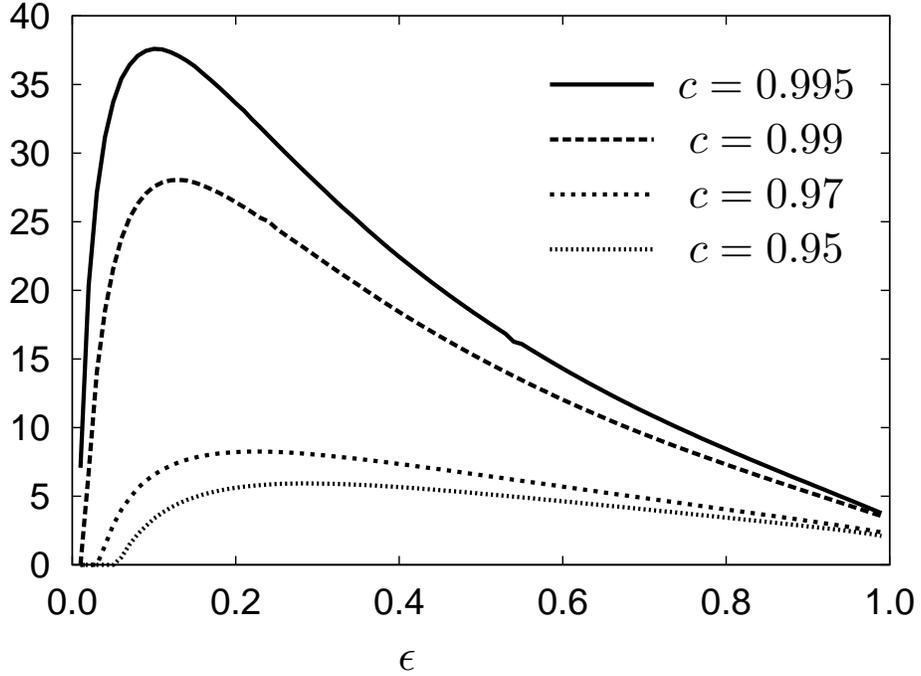}
\caption{\label{FigCutsKh}
Same as Fig.\ref{FigCutsMT}, but for the hard photon emission $K_h$ factor
(see Eqs. (\ref{KDef}), (\ref{eq:hard})).
}
\end{figure}

\begin{figure}
\includegraphics[width=12cm]{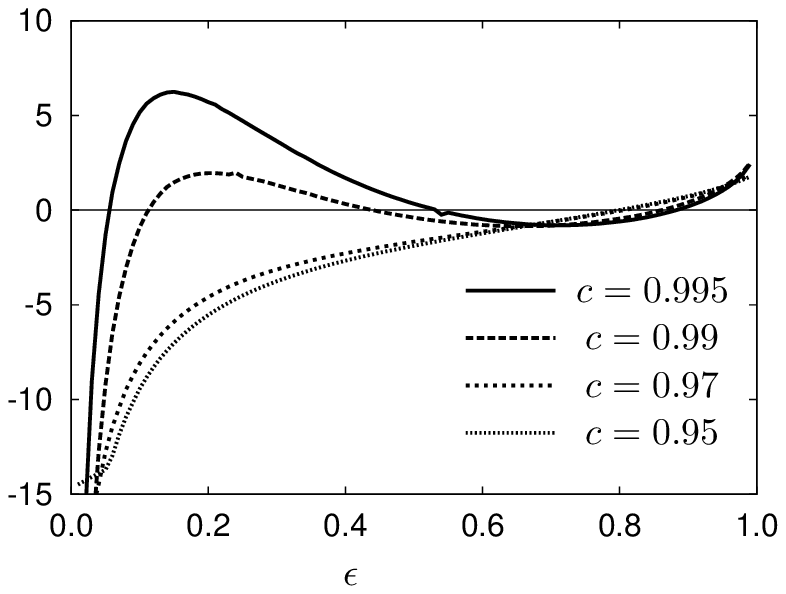}
\caption{\label{FigCutsSum}
Same as Fig.\ref{FigCutsMT}, for the sum $K_o+K_h$.
}
\end{figure}

To demonstrate this cancellation more  we plot in Figs.
Fig.~\ref{FigCutsMT}, \ref{FigCutsKh}, \ref{FigCutsSum} the $\epsilon$-dependence of the $K$-factors and their sum,
for $Q^2 = 3~\GeV^2$ and for different values of the
inelasticity cut $c=0.995, 0.99, 0.97, 0.95$.
In Fig.~\ref{FigCutsSum} the sum of these two contributions is much smaller than
the individual values of the $K$-factors and does not have evident $\epsilon$-dependence.


\begin{figure}
\includegraphics[width=12cm]{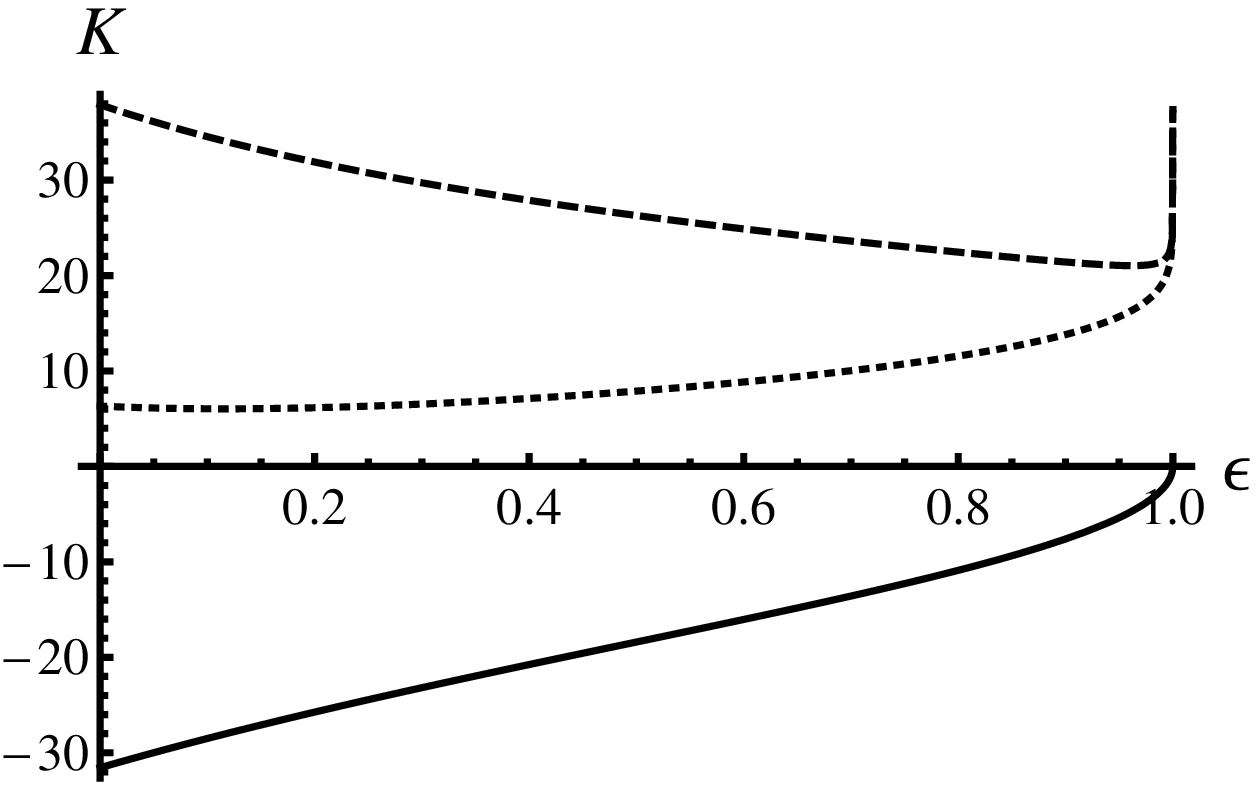}
\caption{\label{FigKWithCut}
Longitudinal and transverse polarized cross sections divided by the Born cross section, for $Q^2$=1,3 and 5 GeV$^2$ (from top to bottom). PL/PL(solid line) all corrections without hard photon,
dashed line: PT
}
\end{figure}

The $K$ factor is shown in Fig. \ref{FigKWithCut} for Q$^2=3~\GeV^2$. $K_h$ (dashed line) compensates essentially the $K_o$ contribution, and in particular flattens the slope of the $\epsilon$ dependence of this contribution.

This supports the calculation and the conclusions of Ref. \cite{Bystritskiy:2007hw}, as the missing contributions essentially cancel.

The results showed that
radiative corrections are driving the slope of the Rosenbluth plot, and, hence, the
extraction of the electric (and magnetic) form factor. Different calculations may give different slopes, and bring the unpolarized and polarized experiments into agreement, without advocating a large contribution of the two-photon exchange diagram.
\section{Application to experiment}
Although it makes sense to compare different calculations, in the same kinematical regions, one should be more careful, indeed, in the comparison with experiment, as RC corrections have to be convoluted with other corrections, such as acceptance and background subtraction.

However, we give here two examples, where the comparison between our calculation and experimental data seems meaningful to us.

\subsection{Unpolarized cross section data}

In single arm experiments, RC depend on an inelasticity cut, which can be done over the outgoing electron energy spectrum or
over the missing mass spectrum. In Ref. \cite{Bystritskiy:2007hw}, an average inelasticity cut of $\Delta E/E_e= 3\%$  was taken. We recalculate here
the corrections $\delta$ for the data \cite{Andivahis:1994rq} using the kinematics and the cuts point by point as in Table \ref{table:table1}.

\begin{figure}
\includegraphics[width=12cm]{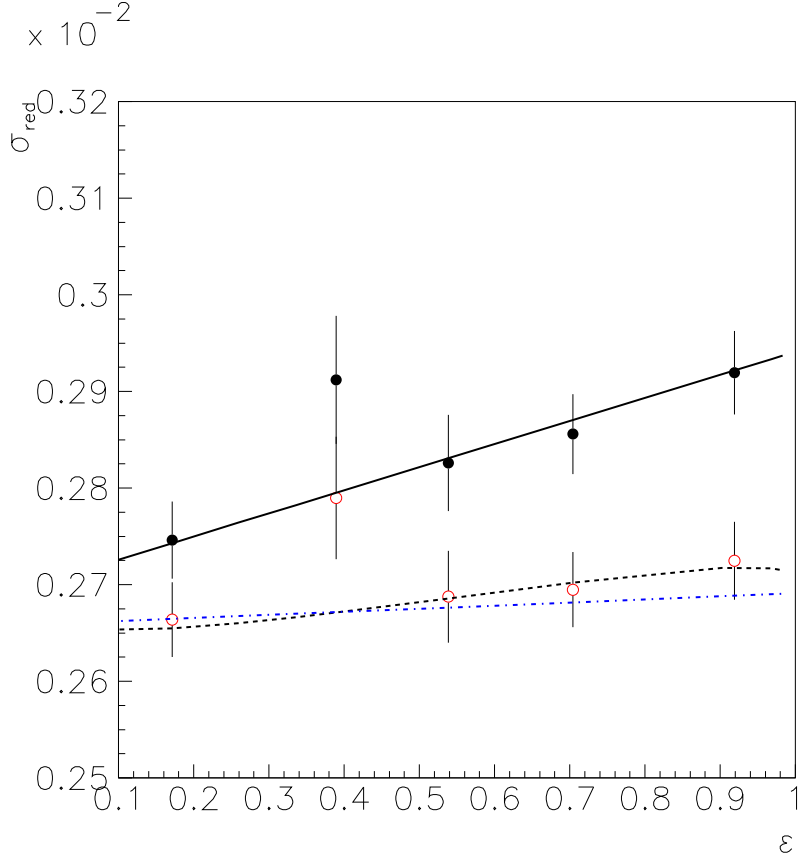}
\caption{\label{Fig:cfp} $\epsilon$-dependence of the reduced cross section at $Q^2=5~\GeV^2$: data as published in from
\protect\cite{Andivahis:1994rq} (black solid points), data corrected by the SF method from Ref. \protect\cite{Bystritskiy:2007hw}. Also shown are the predictions (slopes) from dipole parametrization (solid black line), slope from  SF calculation Ref. \protect\cite{Bystritskiy:2007hw} (dashed black line), slope expected from polarization measurements (dot dashed blue line).
}
\end{figure}

The agreement between the data corrected according to the calculation of Ref. \cite{Bystritskiy:2007hw} becomes satisfactory
(see Fig. \ref{Fig:cfp}).
\begin{center}
\begin{table}[ht]
\begin{tabular}{|c|c|c|c|c|c|c|c|c|}
\hline
$E_e$ [$\GeV$]	&  $\epsilon$ &	   $\theta$  &	 $E'$ [$\GeV$] & Cut $(M_W)$& $\Delta E/E_e$ & $ \delta$ \protect\cite{Andivahis:1994rq} &$ \delta (LSF)$\\
\hline\hline
3.4  & 0.17 & 89.98 &  0.73550 & 1.072 & 0.030& 0.841 & 0.971 \\
\hline
3.96 & 0.39 & 59.29 &  1.29150 & 1.103 & 0.030 & 0.811 & 0.958\\
\hline
4.51 & 0.54 & 45.66 &  1.84250 & 1.131 & 0.030 & 0.801 & 0.951\\
\hline
5.51 & 0.70 & 32.83 &  2.84250 & 1.150 & 0.038 & 0.779 & 0.944\\
\hline
9.80 & 0.92 & 17.523 &  7.13550 & 1.146 & 0.014 & 0.713& 0.933\\
\hline\hline
\end{tabular}
\caption[]{Kinematical table corresponding to the experiment in Ref. \protect\cite{Andivahis:1994rq}: incident energy, $E_e$, $\epsilon$, electron scattering angle ($\theta$) and energy ($E'$), cut on the missing mass, corresponding energy cut, radiative correction $\delta$ from LSF calculation.}
\label{table:table1}
\end{table}
\end{center}

\subsection{Polarization observables}

In Ref. \cite{Meziane:2010xc} the $\epsilon$ dependence of the polarization ratio was measured, at a value of $Q^2=2.49~\GeV^2$. The polarization ratio is related to the ratio $P_T/P_L$, the transverse and longitudinal polarizations of the recoil proton \cite{Akhiezer:1968ek,Akhiezer:1974em}.
 A constant value is predicted within the Born approximation. Corrections beyond the Born approximation would introduce deviations from a constant as well as non-linearities, their size depending on the model. The published experimental values are not corrected by radiative corrections, as they cancel in the ratio (at least at first order, or when can be factorized). However the separate values of $P_L$ and $P_T$ are related to polarized cross section and deviate from Born approximation, due to RC. In polarization experiments, the electron is detected in a calorimeter, therefore all energy is integrated and it is not possible to disentangle the contribution of collinear photon emission along the final electron.

\begin{figure}
\includegraphics[width=12cm]{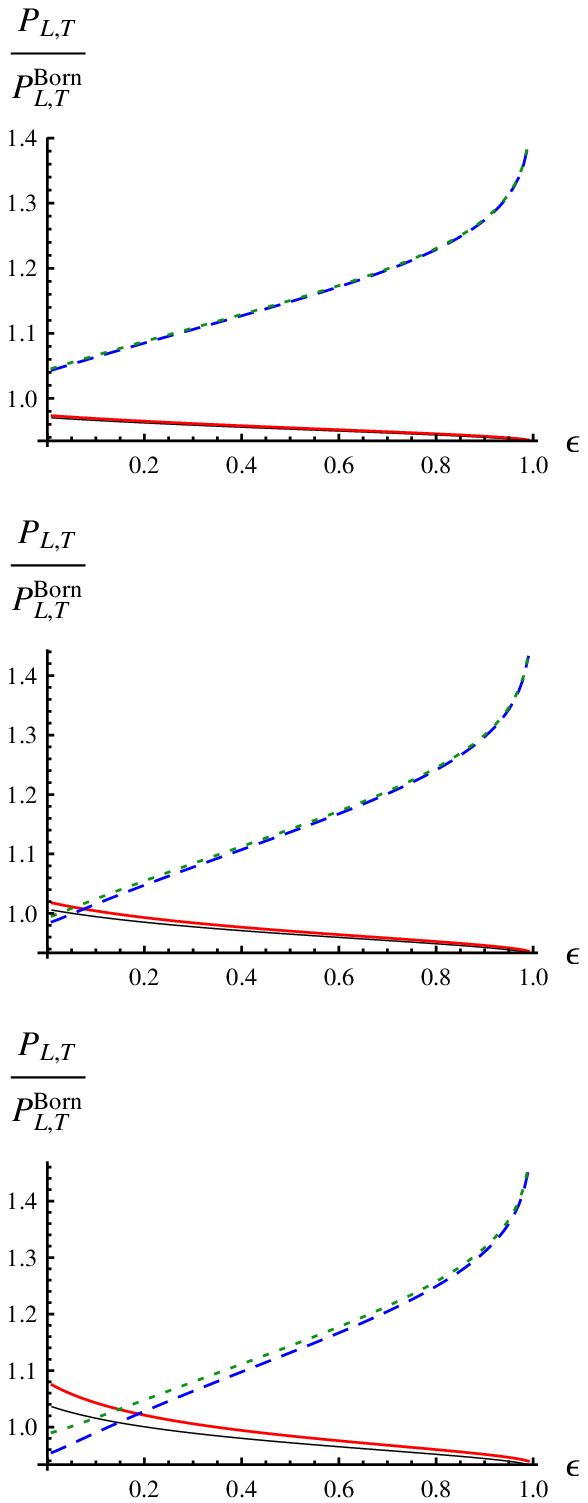}
\caption{\label{PLPTWithHardPhoton}
Longitudinal and transverse polarized cross sections divided by the Born cross section,
for $Q^2$=1, 3 and 5 GeV$^2$ (from top to bottom). The notation is as follows:
$PL/PL^{Born}$ before(black, thin line)   and after (Blue, Dashed line) including hard photon emission, $PT/PT^{Born}$ before (Red, Thick line)  and after (Green Dotted) including hard photon emission
).
}
\end{figure}

Let us apply the corrections from lepton structure functions (LSF), and the present calculation, which includes soft and hard photon emission from final electron. The results, taking into account the hard photon contribution, are shown in Fig. \ref{PLPTWithHardPhoton}. where the ratio of the corrected to Born polarized cross section is plotted for $Q^2$ =1, 3, 5 GeV$^2$, $\Delta_E=0.03$ and $0.99\le c_{3}\le 1$.
As already shown in Ref. \cite{Bystritskiy:2007hw} paper, one can see that the corrections on $P_L$ and $P_T$ may be large and are similar, therefore cancelling in the ratio.

Comparing with Fig. 6 of \cite{Bystritskiy:2007hw} paper,
one can see also that adding the hard photon contribution may change the trend of the curves.
The slope and the magnitude of the correction strongly depend of the energy and angular cuts. Therefore it would be unfair to compare directly with the experimental points from \cite{Meziane:2010xc}, although one can see that the trend of the experimental points can be recovered.

Note that, due to the fact that in the ratio $P_T/P_L$ RC essentially cancel, and therefore non-linearities mostly vanish, the calculation based on LSF reproduces very well this ratio, as already pointed out
in Ref. \cite{Meziane:2010xc}.

\section{Conclusions}

We have calculated radiative corrections for electron--hadron elastic scattering, in frames of the LSF method, and compared to lowest order calculations. We can draw the following conclusions.

Radiative corrections by the LSF method are in general of the same sign, negative, but smaller than for MT, and they have the effect to increase the cross section, when compared to the calculations at the lowest order. The two calculations should basically agree at the lowest order of PT. The difference between the two calculations comes mainly from the fact that, in this application of LSF approach, the contribution of the final emission is unity and that higher order are taken into account (in the leading logarithm approximation).

The different sign of the slope of RC as a function of $\epsilon$ comes, in the MT calculation, from corrections due to the box diagram. We remind that, in this calculation, one photon is soft and the other hard, and the corresponding terms are introduced in order to compensate the infrared divergence due to soft photon emission. Indeed, it was shown in \cite{Kuraev:2006ys} in an exact QED calculation for $e\mu$ scattering, that the box contribution is very small. On the other hand, the charge asymmetry, in the reaction $e^+ +e^-\to \mu^+ +\mu^-$ can be measurable, of the order of percent, due to the dominating contribution of soft photon emission.

We have calculated hard photon emission, which is in general neglected when applying RC to experimental data. Such contribution essentially cancel the interference term from soft photon emission, which is included as a $K$ factor in LSF method.

As for the data of Rosenbluth method for larger values of $Q^2$
the slope of $\epsilon$-dependence of reduced cross section become very small
(or even negative) and whole contribution to it is provided by RC and
not by electric form factor $G_E$. In the polarization transfer approach RC related to longitudinal ($\sigma_L$) and transverse ($\sigma_T$)
polarized recoil proton, essentially cancel in the ratio $\sigma_L/\sigma_T$.

This makes the extraction of the ratio $G_E/G_M$ obtained with the polarization transfer method, more reliable at larger transfer momentum squared. However, when RC are applied to $\sigma_L$ and $\sigma_T$, separately, they are as large as the corrections to the unpolarized cross section, and contain a dependence on $Q^2$ and $\epsilon$.

The experimental conditions are taken into account in the present work, by an 'inelasticity cut'.  Whereas this is convenient for comparison with Rosenbluth data which are basically one arm experiment, and the coincidence with the proton detection is mainly used to lower the background, coincidence experiments require corrections embedded in Monte--Carlo simulations which take into accounts cuts in the multidimensional kinematical space (2-dim for a two body process and 5-dim for a 3-body process) and are convoluted with all other kind of corrections, related to background and detector acceptance. That's why we limited our comparison to the $\epsilon$-dependence of two calculations, in the whole kinematical region (the data from unpolarized cross section and the data from polarization experiments), warning that all contributions should be embedded in the analysis program.

\begin{acknowledgments}
Two of us (E.A.K. and Yu.M.B.) acknowledge the RFBR grant 11-02-00112-a for financial support.
The work of Yu.M.B. was partially supported by JINR grant No. 13-302-04.
We thank Professor V.~S.~Fadin and Dr. A.~V.~Gramolin for critical comments and discussion during the workshop 'Scattering and annihilation electromagnetic processes', Trento, February 18--22, 2013.
\end{acknowledgments}


\end{document}